\documentclass[doublecol]{epl2}
\usepackage{graphicx}

\def\beq{\begin{equation}}
\def\eeq{\end{equation}}
\def\beqa{\begin{eqnarray}}
\def\eeqa{\end{eqnarray}}

\def\cH{{\mathcal H}}
\def\cL{{\mathcal L}}

\def\etal{{\sl et al.},}

\def\nonum{\nonumber \\}

\def\nonum{ \nonumber \\}

\def\probe{\mathrm{probe}}

\title{Information compressibility, entropy variation and approach to steady state in open systems}
\shorttitle{Information compressibility} 
\author{Massimiliano Di Ventra and Yonatan Dubi}
\institute{Department of Physics, University of California - San Diego, La Jolla, California 92093-0319, USA }
 \abstract{We introduce the concept of {\em information compressibility}, $K_I$, which measures the relative
change of number of available microstates of an open system in response to an energy variation. We then prove that at the time in
which the system reaches a steady state, the second and third time derivatives of the information entropy are proportional to the
corresponding time derivatives of the energy, the proportionality constant being $K_I$. We argue that if two steady states with
different but same-sign $K_I$ are dynamically connected in a non-adiabatic way it takes a longer time to reach the state with
compressibility closer to zero than the reverse. We also show analytically that for a two-level system in contact with external baths, the information
compressibility is inversely proportional to the temperature measured at any given time by a probe that is
coupled to the system, and whose temperature is adjusted so that the system dynamics is minimally perturbed. This concept, that applies to both classical and quantum open systems, thus
provides insight into the properties of non-equilibrium steady states. }\begin{document}
 \maketitle

\section{Introduction}
Dynamical systems are to some extent always in interaction with one or more external environments. The latter ones may
exchange particles and/or energy with the system of interest. Under certain conditions this leads to a non-equilibrium steady
state (NESS) of the system dynamics. This situation arises in various physical systems, from chemical and biological processes
\cite{Qian} to nanoscale systems \cite{Gasp1}. The properties of NESSs, whether quantum or classical, have long been the subject
of numerous studies but a comprehensive theoretical description is still lacking, especially for NESSs far from equilibrium
\cite{Book}. One aspect, in particular, that has received much less attention is the approach to steady state,~\cite{DiVentra1}
and the entropy variation in such instance.

The difficulty in describing properties of NESSs and the approach to steady state can be in part attributed to the lack of a
general physical quantity that characterizes such states, especially far from thermodynamic equilibrium. Such a quantity needs to
take into account the microscopic dynamics of the system and, at the same time, provide a global physical description that is, in
principle, easy to access theoretically and/or experimentally.

In this Letter we introduce such a quantity that we name {\it information compressibility}, $K_I$. The latter measures the
``easiness,'' or ``difficulty,'' to vary the relative number of available microstates of an open system when its energy changes
due
to the interaction with the environment(s). 
This concept is the information counterpart of the similar concept for solids, liquids or gases, where the standard
compressibility quantifies the relative change of the volume with respect to a pressure variation.

The advantage of the information compressibility in characterizing non-equilibrium systems stems from the fact that it allows us
to establish several results regarding the approach to steady state. We prove analytically that at the time in which the system
reaches a steady state the second and third time derivatives of the information entropy are proportional to the corresponding
time derivatives of the energy. In both cases, the proportionality constant is $K_I$ at that instant of time. The entropy
production at that moment can be thus minimal or maximal according to the sign of the information compressibility and the
concavity or convexity of the energy function. In addition, we can argue about the time it takes the system to reach a steady
state with a given information compressibility from another steady state with different $K_I$, when the two are connected
non-adiabatically. We find that when the two compressibilities have the same sign it takes a longer time to reach the state with
compressibility closer to zero than the reverse process. The reason for this behavior can be intuitively attributed to the fact
that the fluctuations induced by the presence of the environment act against the first process, while they ``help'' the second
one. No such conclusion can be reached if the two states have compressibilities of different sign. We will illustrate these
findings with a model of a two-level system in contact with two thermal baths. In this case, we can also show analytically
another interesting result: the information compressibility is inversely proportional to the temperature measured at any given
time by a probe that is coupled to the system, and whose temperature is adjusted so that the system dynamics is minimally
perturbed. While we cannot prove this statement in the general case, it shows that this concept, which applies to both quantum
and classical open systems, provides a lot of insight into the properties of non-equilibrium steady states.

\section{Definition of information compressibility} By varying the energy of an open system the number of microstates (in the
appropriate phase space) available to it generally changes. This means that there is a loss -- or gain -- of information when the
system interacts with the environment(s). By energy and entropy we mean the following.

Consider the many-body Hamiltonian
\begin{equation}
\hat H_{tot}=\hat H_S+\hat H_B+\hat H_{int},
\end{equation}
where $\hat H_S$ is the Hamiltonian of the system we are interested in, $\hat H_B$ is the Hamiltonian of the bath(s) degrees of
freedom, and $\hat H_{int}$ describes their mutual interaction. The above Hamiltonians may describe either a quantum or a
classical system. To be specific, from now on we refer to the quantum case only, but all considerations we make apply also to the
classical case, with the appropriate change of quantities (e.g., the replacement of the density matrix $\hat \rho$ with the
classical phase density).

Call $\hat \rho_{tot}$ the density matrix associated with the total Hamiltonian.
 We may then define the reduced density matrix
$\hat \rho$ of the system alone by tracing out the degrees of freedom of the bath(s), namely
\begin{equation}
\hat \rho =\textrm{Tr}_B \{\hat \rho_{tot}\}.
\end{equation}
Note that, in general, there is no closed equation of motion for $\hat \rho$. At this stage, this is of no concern to us and we
assume that we know this operator at any given time. We can then define the average energy of the system as
\begin{equation}
E(t)=\textrm{Tr}\{\hat \rho(t) \hat H_S\},\label{energy}
\end{equation}
and the information entropy
\begin{equation}
S(t)=-k_B \textrm {Tr}\{ \hat \rho(t)\ln \hat \rho(t)\}\label{qentropy},
\end{equation}
with $k_B$ the Boltzmann's constant.

Call $\Omega$ the number of microstates available to the system at any given time. In analogy with the standard compressibility
of matter we can thus define the {\em information compressibility} as
\begin{equation}
K_I(t)=\left . \frac{1}{\Omega}\frac{\delta \Omega}{\delta E}\right|_{E(t)},\label{info1}
\end{equation}
namely the relative variation of the number of microstates with respect to an energy variation, evaluated at the instantaneous
energy.

This definition, however, would require an explicit calculation of the number of available microstates. This is not always easy
to do. We thus seek an alternative definition that is computationally more convenient. We note first that in micro-canonical
equilibrium the number of available microstates is related to the entropy $S$ of the system via the relation $\Omega=
\exp(S/k_B)$. If we introduce such definition into Eq.~(\ref{info1}) we get
\begin{equation}
K_I(t)=\frac{1}{k_B}\left . \frac{\delta S}{\delta E}\right|_{E(t)}.\label{info2}
\end{equation}
We could choose Eq.~(\ref{info2}) as definition of information
compressibility in the general case as well, with the information
entropy given by Eq.~(\ref{qentropy}). Note that despite its
similarity with the well-known equilibrium quantity $\beta =1/k_B
T$, with $T$ the temperature, Eq.~(\ref{info2}) depends on time, and
therefore it cannot be generally interpreted as an inverse thermal
energy (except at canonical equilibrium, or for the two-level system
example we give below). Furthermore, a simple functional dependence
of the entropy on energy may not generally exist out of equilibrium
and, if it does, it may not be unique. In fact, since the entropy is
a non-linear functional of the density matrix, while the energy is a
linear one, Eq.~(\ref{info2}) may even provide multiple values at
specific times. Therefore, definition~(\ref{info2}) is of limited
use, except for specific cases. Instead, both the information
entropy and the energy are generally accessible at any given time,
which implies that their time derivatives are known functions. We
thus replace the definition~(\ref{info2}) with the following --
computationally more convenient -- form
\begin{equation}
K_I(t)\equiv \frac{1}{k_B}\left . \frac{\partial S}{\partial t'}\frac{\partial t'}{\partial E}\right|_{t'=t},\label{info3}
\end{equation}
with $S(t)$ and $E(t)$ evaluated as in Eqs.~(\ref{qentropy}) and~(\ref{energy}), respectively. We stress that although
Eqs.~(\ref{info1}),~(\ref{info2}), and~(\ref{info3}) could all be equally used to define a measure of information
compressibility, we will show in the subsequent discussion that the last one is a natural extension of the concept of inverse
temperature out of equilibrium. Therefore, this is the definition of information compressibility we postulate from now on.

Before discussing the open-system problem let us check the physical
meaning of Eq.~(\ref{info3}) for known cases. First of all, we
realize that if the system is in a steady state the energy variation
is zero, as well as the entropy variation. The information
compressibility would thus acquire a constant value,\footnote{In
some cases $K_I$ may even diverge, indicating that energy relaxes
faster than the entropy.}  which depends on the {\em dynamics} of
the system approaching the given steady state. In addition, the
information compressibility~(\ref{info3}) may be either positive or
negative at any given time, with the negative sign indicating an
increase (decrease) of entropy with decreasing (increasing) energy
supplied to the system by the environment(s).

For a closed system the information entropy does not vary in time, {\em irrespective} of the energy variation \cite{Balian}.
Therefore, as expected, the information compressibility for closed systems is zero at any given time: no matter how much the
energy of a closed system varies, the number of available microstates is constant in time (property of the unitary evolution).

Let us now consider a system in global canonical equilibrium with a bath at temperature $T$. In this case the entropy is $ S_C=
k_B\ln {\cal Z}_C + E_{eq}/T$, where $E_{eq}=\textrm{Tr}\{\hat \rho^{eq}_C \hat H_S\}$ is the average energy of the system at
equilibrium, ${\cal Z}_C$ is the canonical partition function, and $\hat \rho^{eq}_C$ the canonical density matrix. By
differentiating this entropy with respect to the energy (at fixed volume and number of particles) we get $K_I^{C}=1/k_B T$, which
is the well-known quantity $\beta$ of equilibrium statistical mechanics. \cite{Balian} The information compressibility is thus a
natural extension of $\beta$ to the non-equilibrium case, where the concept of temperature is less obvious.

This result also shows that the higher the temperature the more difficult it is to change the relative volume of available
microstates. In fact, a large temperature means a highly-disordered state, and it is thus natural to think that, for a given
small energy variation, the relative change of entropy (or level of disorder) would be small by adding (or subtracting) more
states. Conversely, a small temperature means more order in the system. A small change in energy thus produces a relatively
larger change in the number of microstates available to the system.

\section{Approach to steady state} To illustrate the usefulness of the concept of information compressibility, let us now
examine the role of $K_I$ in the approach to steady state. If we differentiate Eq.~(\ref{info3}) with respect to time we get
\begin{equation}
k_B\frac{dK_I}{dt}\frac{dE}{dt}=\frac{d^2S}{dt^2}-k_BK_I\frac{d^2E}{dt^2}.\label{info4}
\end{equation}
Let us now assume that the system reaches a steady state during time evolution. Call $t_{ss}$ the time at which this
occurs.~\footnote{Note that this time may be infinite.} At steady state $dK_I/dt=0$ (as well as $d E/dt=0$) so that from
Eq.~(\ref{info4}) we get
\begin{equation}
\left. \frac{d^2S}{dt^2}\right |_{t=t^-_{ss}}=k_B\left.K_I\frac{d^2E}{dt^2}\right |_{t=t^-_{ss}},\label{info5}
\end{equation}
where $t=t^-_{ss}$ means the limit in which the time approaches $t_{ss}$ from the past. We can differentiate once more
Eq.~(\ref{info4}), and at steady state we obtain the relation
\begin{equation}
\left. \frac{d^3S}{dt^3}\right |_{t=t^-_{ss}}=k_B\left.K_I\frac{d^3E}{dt^3}\right |_{t=t^-_{ss}},\label{info6}
\end{equation}
which is an even stronger property than Eq.~(\ref{info5}). However, further differentiation of Eq.~(\ref{info4}) does not provide
a simple relation between the $n$-th time derivative of the entropy and the $n$-th time derivative of the energy, even at steady
state. The reason is that time derivatives of $K_I$ of order higher than one appear in such expression, and these derivatives are
not necessarily zero at $t_{ss}$. The above relations also show that in the case of non-equilibrium steady-states, the entropy
can be either maximal or minimal at the onset of steady state. This depends on both the concavity or convexity of the energy
function at that time -- i.e., whether the energy is supplied to or taken away from the system by the environment(s) -- and the
sign of the information compressibility. These functions, in turn, depend on the dynamics (and initial conditions) of the system
approaching the steady state, with the approach possibly being non-monotonous. Note also that the reverse is not necessarily
true: the validity of Eq.~(\ref{info5}) is not sufficient to guarantee that the system is in a steady state.

Having discussed the entropy variation at steady state, let us now apply the above concept to study the time it takes the system
to go from a given steady state with information compressibility, $K^{(1)}_I$, to a steady state with different (but equal sign)
compressibility, $K^{(2)}_I$, and the reverse process (see inset of Fig.~\ref{fig1}). Let us call $S_1$ and $E_1$, the
information entropy and energy of the first steady state, respectively, and $S_2$ and $E_2$, the corresponding quantities of the
second.

We are concerned here with two different steady states which are connected non-adiabatically, and have the same-sign
$K_I$.~\footnote{``Non-adiabatic'' means that the physical properties of the environment (e.g., its temperature) are changed
faster than the dynamics of the system. An adiabatic path between the two steady states results, by construction, in a
time-symmetric evolution from one to the other, and vice versa.} For clarity, let us assume
\begin{equation}
K^{(1)}_{I}>K^{(2)}_{I},
\end{equation}
and both quantities are positive. Let us first consider the time $t^{(1)}_{ss}$ at which we increase the energy of the system
(which implies that the entropy increases as well) so that the latter evolves from the steady state 1 to the steady state 2
($dK_I/dt<0$ at $t^{(1)}_{ss}$). Referring to Eq.~(\ref{info4}), we then see that its
 lhs is a negative quantity. Since, by assumption, both $d^2
S_1/dt^2$ and $d^2 E_2/dt^2$ are positive at $t^{(1)}_{ss}$, we find that at that instant
\begin{equation}
\frac{d^2 S_1}{dt^2}<k_BK^{(1)}_{I}\frac{d^2 E_1}{dt^2},
\end{equation}
namely the entropy variation is ``slower'' than the corresponding energy variation. If we repeat the same reasoning for the time
$t^{(2)}_{ss}$ at which we {\em reduce} the energy (and entropy) to recover steady state 1 from steady state 2 (we are assuming
here that all these times are finite), we find instead
\begin{equation}
\left|\frac{d^2 S_2}{dt^2}\right |>k_BK^{(2)}_{I}\left|\frac{d^2 E_2}{dt^2}\right |,
\end{equation}
namely the disorder varies faster than the corresponding energy variation. Note that these arguments depend only on the
properties of the information compressibility close to the initial and final state, irrespective of the full dynamics. We thus
argue that it takes {\em longer} to reach the steady state with compressibility closer to zero than the reverse process. The same
can be concluded for two steady states with negative compressibilities. This reflects the intuitive notion that the fluctuations
induced by the presence of the environment act against the ``compression'' of the volume of available microstates, while they
favor the reverse process.

Note that one cannot say anything about these time scales when the compressibilities of the two steady states differ in sign. In
such a case, applying the above argument leads to the obvious mathematical statement that a negative number is smaller than a
positive one, thus precluding any conclusion on the time scales of the two processes.

\section{Example} We now illustrate the above results for a specific model system. We study a quantum two-level
system in contact with two heat baths kept at two different temperatures. We work in the Markovian approximation and use the
Lindblad equation ($\hbar=1$) \cite{Lindblad,VanKampen},
 \beqa
\dot{\rho}&=&-i [\cH,\rho]+\cL_1 \rho+\cL_2 \rho \nonum \cL_i&=&-\frac{1}{2}\{V^\dagger_i V_i,\rho\} +V_i \rho V^\dagger_i,
~~i=1,2 \label{Lindblad} \eeqa where $\rho=\left(
              \begin{array}{cc}
                \rho_{11} & \rho_{12} \\
                \rho^*_{12} & 1-\rho_{11} \\
              \end{array}
            \right)$
is the density matrix, and the Hamiltonian and relaxation operators are given by \cite{VanKampen} \beqa \cH&=&\left(
      \begin{array}{cc}
        \omega & 0 \\
        0 & 0 \\
      \end{array}
    \right) \nonum
V_i&=&\gamma^{1/2}_i \left(
        \begin{array}{cc}
          0 & \exp\left( -\frac{\omega}{4 k_B T_i} \right) \\
          \exp\left( \frac{\omega}{4 k_B T_i} \right) & 0 \\
        \end{array}
      \right)\label{hamiltonian}
\eeqa Here $T_i$ ($i=1,2$) is the temperature of the $i$-th bath, $\omega$ is the two-level energy separation, and
$\gamma_i=\gamma_0 \cosh \left( \frac{\omega}{2 k_B T_i} \right)$ is a normalization factor with $\gamma_0$ some relaxation rate.
This form is chosen such that detailed balance is satisfied for each thermal bath independently, and the relaxation rates remain
finite even in the limit $T \to 0$. (One can, in fact, choose different normalizations for the Lindblad operators. Such a change
does not affect qualitatively the results presented here.)

\begin{figure}
\centering \vskip 0.5truecm
\includegraphics[width=8truecm]{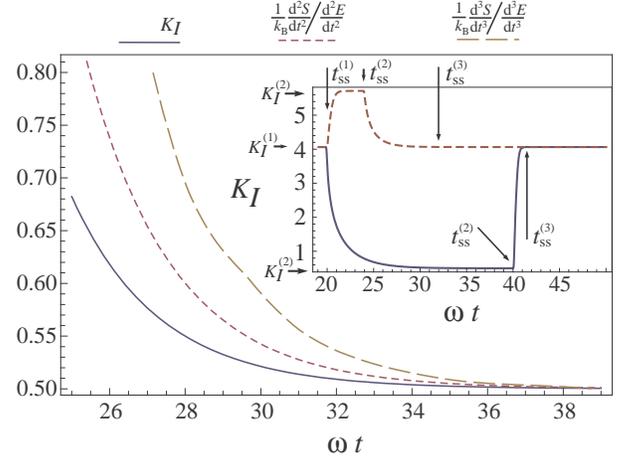}
\caption{\label{fig1} (Color online) The information compressibility $K_I$, and the derivative ratios $\frac{1}{k_B}\frac{d^2
S}{dt^2}/\frac{d^2 E}{dt^2}$ and $\frac{1}{k_B}\frac{d^3 S}{dt^3}/\frac{d^3 E}{dt^3}$ as a function of time (in units of
$1/\omega$) for the example given by Eqs.~(\ref{Lindblad}-\ref{hamiltonian}) (see text for parameters). As the system approaches
the steady state, all three quantities converge into a single value. Inset: dynamics of the information compressibility. As
discussed in the text, when $|K^{(1)}_{I}|>|K^{(2)}_{I}|$, $(t^{(2)}_{ss}-t^{(1)}_{ss})>(t^{(3)}_{ss}-t^{(2)}_{ss})$, while
$|K^{(1)}_{I}|<|K^{(2)}_{I}|$ implies $(t^{(2)}_{ss}-t^{(1)}_{ss})<(t^{(3)}_{ss}-t^{(2)}_{ss})$.}
\end{figure}

The analytical solution of Eq.~(\ref{Lindblad}) for arbitrary times is too cumbersome to be presented here. We thus show only the
main results. We have calculated the dynamics of the system, starting from some initial condition, with the initial parameters
$\omega=1$ (which sets the energy scale), $\gamma_0=0.1$ and $T_1=2, T_2=0.2$. We then let the system evolve into the steady
state. At a time $t^{(1)}_{ss}$ we vary $T_2$ non-adiabatically. We have varied $T_2$ instantaneously both to a value $T_2=2$ (blue solid curve
in inset of Fig.~\ref{fig1}) and to a value $T_2=0.16$ (red dashed curve in inset of Fig.~\ref{fig1}). We again wait until the
system reaches a new steady state. This occurs at a time $t^{(2)}_{ss}$ when we also switch $T_2$ back to its initial value. The
initial steady state is finally reached at a time $t^{(3)}_{ss}$ (see inset of Fig.\ref{fig1}).

In Fig.~\ref{fig1} we plot the information compressibility $K_I$ and the derivative ratios $\frac{1}{k_B}\frac{d^2
S}{dt^2}/\frac{d^2 E}{dt^2}$ and $\frac{1}{k_B}\frac{d^3 S}{dt^3}/\frac{d^3 E}{dt^3}$ as a function of time, for times less than
$t=t^{(2)}_{ss}$ for the case in which $T_2$ has been varied to $T_2=2$. As predicted by Eqs.~(\ref{info5}-\ref{info6}) we find
that the derivative ratios converge to $K_I$ as the system approaches the steady state. We obtain the same result (not shown) for
the time $t^{(3)}_{ss}$, when the system is driven back to the original steady state.

In the inset of Fig.~\ref{fig1} we plot the time-dependence of $K_I$, to illustrate its dynamics for the two cases considered
above. We indeed find that $(t^{(2)}_{ss}-t^{(1)}_{ss})>(t^{(3)}_{ss}-t^{(2)}_{ss})$ when $|K^{(1)}_{I}|>|K^{(2)}_{I}|$ (blue
solid curve) and $(t^{(2)}_{ss}-t^{(1)}_{ss})<(t^{(3)}_{ss}-t^{(2)}_{ss})$ when $|K^{(1)}_{I}|<|K^{(2)}_{I}|$ (red dashed curve),
thus corroborating our analysis.

\section{Local temperature out of equilibrium}
The concept of information compressibility also provides some
insight regarding another interesting question. Consider a system,
like the two-level system discussed previously, to which a
thermometer is (weakly) coupled. In equilibrium, the thermometer
would thus measure the temperature of the system. Now let the latter
evolve. The system then ends up into a new state, which may or may
not be an equilibrium state. Irrespective, during time evolution the
system is out of equilibrium. If we assume that the relaxation time
of the thermometer is faster than the dynamics of the system, then
the thermometer would probe a ``temperature'' that varies in time.
The question is then, what is this time-dependent temperature that
the thermometer is measuring?

We cannot provide an answer to the general case but we can conjecture that under suitable conditions this temperature is
precisely the non-equilibrium generalization of the equilibrium temperature, and is the inverse (in units of $k_B$) of the
information compressibility $K_I$. Let us determine under which conditions this conjecture is correct for the simple two-level
system example. In order to mimic a thermometer connected to the system, we couple the latter to an external probe (which may,
e.g., represent the physical tip of a scanning tunneling microscope-mounted thermometer \cite{Cahill}). Mathematically, this
amounts to adding an additional relaxation operator to Eq.~(\ref{Lindblad}) of the form~(\ref{hamiltonian}), with a temperature
$T_{\probe}$ and some coupling $\gamma_{\probe}$. Now, $T_{\probe}$ is allowed to ``float'', i.e., for any given instant of time
$t$ we vary it until the dynamics of the system is minimally perturbed (we call this a ``floating temperature probe'') \cite{us}.
This is equivalent to assuming that a local equilibrium has been established between the system and the probe at any given time,
again assuming that the relaxation time of the probe is much faster than the system dynamics.

The function $T_{\probe}(t)$ is such that there is no change of
dynamics with or without the probe. Let the density matrix $\rho(t)$
be the solution of Eq.~(\ref{Lindblad}) without the additional probe
operator. Then, in order for the solution not to change in the
presence of the additional probe, the condition
$\cL_{\probe}[T_{\probe}(t)] \rho(t)=0$ must be satisfied. In
general, this condition does not hold unless the off-diagonal
elements of the density matrix are zero at any given time. This
amounts to saying that the environments induce decoherence on time
scales much faster than the time scale over which the system
evolves. Then, using the definition of the relaxation operators in
Eq.~(\ref{hamiltonian}), it is a matter of algebra to find that the
diagonal elements of $\cL_{\probe}[T_{\probe}(t)] \rho(t)$ vanish if
\beq k_B T_{\probe}(t)=\omega \left [\log \left(
\frac{1-\rho_{11}(t)}{\rho_{11}(t)}\right)\right
]^{-1}~~,\label{Ttip}\eeq irrespective of the coupling
$\gamma_{\probe}$.

We now calculate the information compressibility. In this simple example we have $E(t)=\omega \rho_{11}(t)$ and
$S(t)/k_B=-\rho_{11}(t) \log \rho_{11}(t)-(1-\rho_{11}(t))\log (1-\rho_{11}(t))$. Thus, one can explicitly extract $S(E(t))$ and
we have \beq K_I=\frac{1}{k_B}\left . \frac{\delta S}{\delta E}\right|_{E(t)}=\frac{1}{\omega} \log \left(
\frac{1-\rho_{11}(t)}{\rho_{11}(t)}\right)~~ \eeq and hence $K_I(t)=1/k_BT_{\probe}(t)$, as conjectured. We note again that we
cannot give a similar analytic proof for the general case. However, numerical verifications of our conjecture in other systems
are currently underway.

 \section{Summary} In this paper we have introduced the notion of
 \emph{information compressibility} $K_I$ (Eq.~\ref{info3}), which describes the ability of
 an open system to change its entropy in response to a variation of its energy. It is a natural extension of the
 equilibrium thermodynamics quantity $\beta=1/k_B T$ to which it reduces in canonical
 equilibrium.
 This concept, however, extends also far from
 equilibrium, where the notion of temperature is less obvious. In fact, we have analytically shown that for a two-level system coupled to external baths, the
 information compressibility is inversely proportional to the temperature a probe coupled to the system would measure when it minimally perturbs the
 system's dynamics. We could not however prove this result in the general case. We have also
 demonstrated that knowledge of $K_I$ provides information on the
 dynamics of the system as it approaches a steady state, and
 validated our results with a simple example.
 The concepts and results described here are, in principle,
 testable within present experimental model systems \cite{exp}, and we hope this
  work will motivate studies in this direction.

 Finally, we believe that the concept of information compressibility can be a valuable tool in studying
 the dynamics of out-of-equilibrium phenomena, and may be useful in
 characterizing, for instance, dynamical phase transitions such as
 the glass transition, where the properties of the system vary
 dramatically.

 \acknowledgments
 We thank R. D'Agosta, M. Zwolak, and J. Hirsch for useful comments on the manuscript. This work has been funded by the DOE
grant DE-FG02-05ER46204.

\end{document}